\documentclass[10pt,aps,twocolumn,showpacs,floatfix,prb,superscriptaddress]{revtex4-1}

\usepackage{graphicx}
\usepackage{amsmath}
\usepackage{amssymb}
\usepackage{dsfont}
\usepackage{txfonts}
\usepackage[colorlinks,citecolor=blue,linkcolor=blue,urlcolor=blue,plainpages=false,pdfpagelabels]{hyperref}
\usepackage{soul}

\begin{document}

\title{\texorpdfstring{$\mathbb{Z}_4$}{Z4} parafermions in one-dimensional fermionic lattices}

\author{Alessio Calzona}
\email{calzona@fisica.unige.it}
\affiliation{Physics and Materials Science Research Unit, University of Luxembourg, L-1511 Luxembourg}
\affiliation{Dipartimento di Fisica, Universit\`a di Genova and SPIN-CNR, Via Dodecaneso 33, 16146, Genova, Italy}

\author{Tobias Meng}
\affiliation{Institut f\"ur Theoretische Physik, Technische Universit\"at Dresden, 01062 Dresden, Germany}

\author{Maura Sassetti}
\affiliation{Dipartimento di Fisica, Universit\`a di Genova and SPIN-CNR, Via Dodecaneso 33, 16146, Genova, Italy}

\author{Thomas L. Schmidt}
\affiliation{Physics and Materials Science Research Unit, University of Luxembourg, L-1511 Luxembourg}

\date{\today}

\begin{abstract}
Parafermions are emergent excitations which generalize Majorana fermions and are potentially relevant to topological quantum computation. Using the concept of Fock parafermions, we present a mapping between lattice $\mathbb{Z}_4$-parafermions and lattice spin-$1/2$ fermions which preserves the locality of operators with $\mathbb{Z}_4$ symmetry. Based on this mapping, we construct an exactly solvable, local and interacting one-dimensional fermionic Hamiltonian which hosts zero-energy modes obeying parafermionic algebra. We numerically show that this parafermionic phase remains stable in a wide range of parameters, and discuss its signatures in the fermionic spectral function.
\end{abstract}

\maketitle

{\it Introduction.} In its early years, the field of topological states of matter has mainly been centered around non-interacting Hamiltonians and the topology of their band structures \cite{hasan10,qi11,leijnse12,beenakker13}. In electronic systems, however, the presence of Coulomb repulsion raises the question to which degree topology and interactions coexist or compete. It has by now become clear that there is no general answer to this question -- the effect of interactions can range from perturbatively small modifications of effective band structures to a complete loss of the topological distinction between different phases. As a third and much more exciting option, interactions can give rise to entirely new phases of matter, a prime example of which are topologically ordered states such as fractional quantum Hall states. These systems feature emergent low-energy excitations called anyons that behave differently from fermions or bosons \cite{nayak08}. Most sought-after are non-Abelian anyons in many-particle systems with a topologically protected ground state degeneracy. Braiding two of these non-Abelian anyons implements a rotation in the degenerate ground state manifold, which in turn allows to perform quantum computation in a way that minimizes decoherence at the hardware level \cite{Kitaev03,Aasen16,nayak08}.
	
As a major breakthrough, it has been realized that anyonic excitations can not only exist as quasiparticles of strongly interacting systems such as fractional quantum Hall states, but may also emerge as special bound states of quadratic Hamiltonians. The best-studied example here are Majorana fermions, which appear as vortex-bound states in $p$-wave superconductors \cite{volovik_book}, and at domain walls of simple chains of superconducting spinless electrons \cite{kitaev01,alicea12}. A large body of research has been devoted to the experimental realization of Majorana bound states, including in particular semiconducting quantum wires with strong spin-orbit coupling \cite{mourik12,albrecht16,Lutchyn17,Oreg10,Lutchyn10,Deng16} or magnetic adatoms on superconductors \cite{nadjperge14,Choy11,Nadj13,Klinovaja13rkky,Ruby15,Feldman17}. While Majorana fermions are the simplest example of non-Abelian anyons, they are not complex enough to implement universal quantum computing. More recently, the focus has thus shifted to so-called $\mathds{Z}_n$-parafermions, generalizations of Majorana fermions associated with richer braiding properties \cite{nayak08,Hutter16}.

These more complicated parafermions cannot be realized in non-interacting Hamiltonians, but rather are an example of a topological phenomenon that only exists in the presence of electron-electron interactions. Various experimental realizations for some of those parafermions have been proposed, for instance quantum spin Hall systems \cite{orth15,vinkleraviv17,zhang14,klinovaja14,Fleckenstein2018}, quantum wires \cite{pedder16,pedder17,klinovaja13,klinovaja14}, fractional quantum Hall insulators \cite{barkeshli14,clarke13,lindner12,Alavirad17,Chen16,Vaezi13,Vaezi14,Vaezi17} and other systems \cite{barkeshli13,santos17}. These theoretical studies, however, are all based on effective low-energy field theories and extensions of Luttinger liquid physics. Hence, the parafermions they predict are not exact eigenstates of a microscopic fermionic Hamiltonian. A complementary line of research has been devoted to more mathematical studies of intrinsic properties of parafermions chains \cite{iemini17,mazza18,jermyn14,fendley12,fendley14,alicea16,Mong14,Stoudenmire15,Meichanetzidis17,Mazza2018}. In the present Rapid Communication, we propose for the first time an exact mapping between parafermionic chains and 
electronic Hamiltonians on a lattice. This mapping not only provides an insightful bridge between mathematical models and physical systems, but also paves the way for the systematic implementation and analysis of parafermionic Hamiltonians using fermions.

{\it From parafermions to fermions.} The starting point of our analysis is a one-dimensional chain of $\mathbb{Z}_4$-parafermions, which is a natural generalization of the Kitaev chain of Majorana fermions. At each site $i \in \{1, \ldots, L\}$, we define two parafermionic operators $a_i$ and $b_i$, which satisfy $a_i^4 = 1$ and $a_i^{3} = a_i^\dagger$ (the same applies for $b_i$). By definition, parafermionic operators satisfy the anyonic exchange statistic: $a_l a_j = i\, a_j a_l, \; b_lb_j = i \, b_jb_l$ for $l<j$, and $a_l b_j = i\, b_j a_l $ for $ l\leq j$.
To relate these operators to physical electrons, we study how they act on the states of the system. The fact that $a_i^4 = 1=b_i^4$ implies that each lattice site can be associated with four different states, and that the application of the operators $a_i$ and $b_i$ cycles through those states. This notion can be made more precise by associating a Fock space to the parafermionic operators via the introduction of ``Fock parafermions'' (FPF) \cite{cobanera14,supp}. The latter are described by creation ($d^\dagger_j$) and annihilation ($d_j$) operators, which allow us to express $a_i$ and $b_i$ as
 \begin{align}
  a_j = d_j + d_j^{\dagger 3}, \quad
  b_j = e^{i \pi/4} (d_j i^{N_j} + d_j^{\dagger 3}) ,
 \end{align}
where $N_j= \sum_{m=1}^3 d_j^{\dag m}  d_j^m$ is the number operator for FPF whose four integer eigenvalues run from $0$ to $3$. 
The parafermionic algebra of $a_i$ and $b_i$ is handed down to the FPFs in their commutations relations: $d_ld_j=i\;d_jd_l$ and $d_jd^\dagger_l=i\;d^\dagger_ld_j$ for $ l<j$. Moreover, on a given site one has $d_j^{\dagger m}d_j^m+d_j^{(4-m)}d_j^{\dagger(4-m)}=1$, for $m=1,2,3$, and $d_j^4=0$.

The single-site four-dimensional parafermionic Fock space can be identified with the Fock space of spin-$1/2$ fermions, inducing an on-site mapping between FPF operator and physical fermions. The introduction of appropriate string factors allows to extend the mapping over the whole chain, in analogy with the well-known Jordan-Wigner transformation between spin chains and fermionic ones. Since the identification between the two Fock spaces not unique, one can find many different  mappings between FPFs and fermions on a lattice. Here we consider in particular the following one (derived in Appendix \ref{app:A})
\begin{equation}
\label{eq:mapping_full}
\begin{split}
d_j &= i^{\sum_{p<j} \left(  n_{p\downarrow} + 3n_{p\uparrow} - 2 n_{p\uparrow}n_{p\downarrow} \right)}\;  \big( c_{j\uparrow} - c_{j\uparrow} n_{j\downarrow} -c^\dagger_{j\uparrow} n_{j\downarrow} + i  c_{j\downarrow} n_{j\uparrow} \big) 
\end{split}	
\end{equation}
which features a definite odd fermion parity ($n_{j\sigma} = c^\dag_{j\sigma} c_{j\sigma}$). This property is crucial since it remarkably ensures that every local parafermionic operator which conserves the number of FPFs modulo $4$ is transformed into a fermionic operator \emph{without} string factors (see Appendix \ref{app:A}). Despite the high non-locality of Eq.\ \eqref{eq:mapping_full}, it is therefore possible to map parafermionic nearest-neighbor Hamiltonians into fermionic models with on-site and nearest-neighbor terms only. 

	
{\it Mapping of the Hamiltonian.} In the remainder, we apply the above mapping to the following  $\mathds{Z}_4$-parafermionic Hamiltonian on an open $L$-site chain \cite{alicea16}
	\begin{equation}
	\label{eq:H_pf}
	H_J = - J e^{i \frac{\pi}{4}} \sum_{j=1}^{L-1} b_j a^\dagger_{j+1}\; + \text{h.c.}\,,
	\end{equation}
where we assume $J>0$.
This exactly solvable model \cite{iemini17,supp} can be seen as a generalization of Kitaev's Majorana chain model and has a non-trivial topology. There are two dangling parafermions, $a_1$ and $b_L$, which commute with the Hamiltonian and induce an exact and topologically protected $4$-fold degeneracy throughout the entire spectrum. As one important feature, the Hamiltonian $H_J$ has a $\mathbb{Z}_4$ symmetry $\mathcal{Z}=i^{\sum_j N_j}$ which guarantees the conservation modulo $4$ of the total number of FPFs.
	
The mapping~\eqref{eq:mapping_full} allows us to translate the Hamiltonian~(\ref{eq:H_pf}) to a local fermionic Hamiltonian $H_J = H^{(2)} + H^{(4)} +H^{(6)}$,  with 
			%
\begin{align}
\label{eq:H2}
&H^{(2)} =- J \sum_{\sigma,j} \left[c_{\sigma,j}^\dagger c_{\sigma, j+1} -i \;c_{-\sigma,j}^\dagger c^\dagger_{\sigma,j+1} \right] +h.c.\,,\\
\label{eq:H4}
&H^{(4)}= -J\sum_{\sigma,j} \Big[  c_{\sigma,j}^\dagger c_{\sigma, j+1} \left(- n_{-\sigma,j}-n_{-\sigma,j+1}\right)   \notag \\[-.3em]
& \quad + c_{\sigma,j}^\dagger c_{-\sigma, j+1}\; i \left( n_{-\sigma,j} + n_{\sigma,j+1} \right) +  c_{-\sigma,j}^\dagger c^\dagger_{\sigma,j+1}\; i \left(n_{\sigma,j}+n_{-\sigma,j+1}\right) \notag \\
& \quad + c_{\sigma,j}^\dagger c^\dagger_{\sigma,j+1} \left(n_{-\sigma,j}-n_{-\sigma,j+1}\right)\Big] + h.c.\,, \\
\label{eq:H6}
&H^{(6)}= - J \sum_j \Big[ - 2 i\,   c_{\sigma,j}^\dagger c_{-\sigma, j+1} \left( n_{-\sigma,j} n_{\sigma,j+1}\right)\notag \\
& \qquad \qquad  - 2i \, c_{-\sigma,j}^\dagger c^\dagger_{\sigma,j+1} \left(n_{\sigma,j} n_{-\sigma,j+1}\right) \Big]  +h.c.\,.
\end{align}
In the fermionic language, $H_J$ consists of superconducting pairing and hopping terms with and without spin-flip, locally weighted by the fermion occupation numbers on the lattice sites. Note that the Hamiltonian $H_J$ is time-reversal invariant. 

The mapping we have developed allows to express parafermionic operators in terms of electrons. The zero-energy parafermionic modes, in particular, have the following fermionic expression:
\begin{align}
a_1 &= i c_{1\downarrow} n_{1\uparrow}-c^\dagger_{1\uparrow}n_{1\downarrow}+c_{1\uparrow}(1-n_{1\downarrow})+ic^\dagger_{1\downarrow}(1-n_{1\uparrow}) \notag \\
b_L&= e^{i \pi/4} \; (-i)^{\sum_{j=1}^{L-1} N_j} \; \big[i c^\dagger_{L\uparrow}n_{L\downarrow}+ic_{L\uparrow}(1-n_{L\downarrow})\notag \\&\quad- ic^\dagger_{L\downarrow}(1-n_{L\uparrow}) -i c_{L\downarrow} n_{L\uparrow}
\big]\,.\label{eq:b_L}
\end{align}
These equations represent an important result, namely the explicit expression of combinations of fermionic operators that satisfy the parafermionic algebra and that commute with the fermionic Hamiltonian $H_J$.

An important question concerns the locality and topological protection of the zero-energy states of the fermionic Hamiltonian. Although $a_1$ and $b_L$ are localized at the edge in the parafermionic language, one of the corresponding fermionized operators (in our case $b_L$) inevitably contains a non-local string factor. This string factor is not associated with a density of states (see below), but allows the edge mode to ``feel'' what happens in the bulk. The non-locality hence challenges the \emph{topological} protection of the fourfold ground state degeneracy in the fermionic model. As we will discuss in the next section, it indeed turns out that only a twofold degeneracy is topologically protected. Remarkably, the non-locality of $b_L$ does not prevent us from finding local operators on either edge of the fermionic chain that cycle through the four degenerate ground states (see Appendix  \cite{app:B}).

{\it Topological properties of the fermionic chain.} In the parafermionic language, the model in Eq.~\eqref{eq:H_pf} represents  a topological phase \cite{fendley12,fendley14,alicea16} in which the spectrum exhibits a topologically protected fourfold degeneracy that cannot be lifted by local parafermionic perturbations. It is natural to ask if the same holds also for the corresponding fermionic chain, since it is well known that the presence of string factors can change the topological properties of the system. The prototypical example is the Kitaev chain, which is related by a non-local Jordan-Wigner transformation to the topologically trivial quantum Ising model \cite{kitaev01}. 

{In this respect, it is instructive to study the symmetries featured by the fermionic model in Eq.\ (\ref{eq:H2}-\ref{eq:H6}). The $\mathbb{Z}_4$ symmetry of the parafermionic Hamiltonian in Eq.\ \eqref{eq:H_pf} can be expressed in terms of fermions as $\mathcal{Z} = i^{\sum_j[(n_{j\uparrow}+n_{j\downarrow})^2+2n_{j\downarrow}]}$. Its square corresponds to the usual $\mathbb{Z}_2$ fermion parity $\mathcal{P}=\mathcal{Z}^2 = (-1)^{\sum_j (n_{j\uparrow}+n_{j\downarrow})}$. Interestingly, the local operator $M_j = i \gamma_{\uparrow,j} \gamma_{\downarrow,j}$, where $\gamma_{\sigma,j}=c^\dagger_{\sigma,j}+c_{\sigma,j}$ are Majorana operators, commutes with the Hamiltonian but anticommutes with the  $\mathbb{Z}_4$ symmetry $\{M_j,\mathcal{Z}\}=0$. It can be therefore identified as a $\mathbb{Z}_2$ local order operator, associated with the $\mathbb{Z}_2$ symmetry $\mathcal{S}_B= e^{-i \frac{\pi}{4}}2^{-\frac{1}{2}}\;  \mathcal{Z} + h.c.$  which is spontaneously broken and satisfies $[\mathcal{S}_B,H_J]=\{\mathcal{S}_B, M_j\}=0$.} This local order parameter thus differentiates the four degenerate ground states into two pairs and the degeneracy between them can be split by a local perturbation containing any of the $M_j$. A concrete example of such a perturbation is a magnetic field along the $y$ axis
\begin{equation}
H_{y} = B_y \sum_{i=1}^L i \left(c^\dagger_{j,\uparrow}c_{j,\downarrow}-c^\dagger_{j,\downarrow}c_{j,\uparrow}\right)=B_y \sum_{i=1}^L \tfrac{1}{2} \left(M_j + i \eta_{\uparrow,j}\eta_{\downarrow,j}\right)
\end{equation}
where $\eta_{\sigma,j}=i(c^\dagger_{\sigma,j}-c_{\sigma,j})$ are the other Majorana operators. Our DMRG simulations indeed confirm that even a small field $B_y$ reduces the fourfold degeneracy to a doublet of twofold (almost) degenerate states, with an energy difference which scales linearly with the system size $L$ (see Appendix \ref{app:C}). 

On the other hand, the fourfold degeneracy is protected against other local perturbations, including in particular 
a magnetic field in the $(x,z)$-plane or a chemical potential: our DMRG calculations indicate that the lifting induced by these perturbations is exponentially suppressed in the system length (see Appendix \ref{app:C}). The protection of the degeneracy against some of these perturbations becomes apparent in the parafermionic language. Both the chemical potential and the magnetic field along the $z$ axis conserve the total number of FPF modulo $4$ and they thus feature a local expression also in terms of parafermions (see Appendix \ref{app:C}). 
Our findings are consistent with the results of Ref.~\cite{fidkowski11,Turner11,Bultinck17} in that the fully topologically protected part of the degeneracy (the part that cannot be lifted by symmetry-breaking local perturbations) is only twofold. 

{\it Phase diagram.} 
Being an exactly solvable model, $H_J$ allowed us to derive important analytical results such as the existence of the local order parameter $M_j$ and the expression of the zero-energy parafermions in Eq.~\eqref{eq:b_L}. The price we payed for this exact solvability is the rather complicated form of $H_J$ in the fermionic basis, which in particular includes  three-body interactions. Instead of searching for fine-tuned models that might realize Eq.~(\ref{eq:H2}-\ref{eq:H6}), we view this model as one representative of a much larger class of systems realizing parafermionic physics at low energies. In this spirit, the specific model $H_J$ is not only crucial in that it allows us to fully understand the physics beyond any low-energy approximations, but also as a controlled starting point around which we now explore topologically equivalent models by smooth deformations of the Hamiltonian. As long as the gap is not closed, the system remains in the same topological phase and will feature the same topological properties. In particular, we consider the much more generic Hamiltonian 
\begin{equation}
\bar H (U,V) = H^{(2)} + U \left[ V \left(H^{(4)} + H^{(6)} \right) + (1-V) \bar H^{(4)}\right]
\end{equation}
where the parameter $U$ weights all interacting terms and $V$ allows to smoothly transform the three-body terms into simpler two-body terms with
\begin{equation}
\begin{split}
\bar H^{(4)} = -J \sum_{\sigma,j} \Big[&c_{\sigma,j}^\dagger c_{\sigma, j+1} \left(-n_{-\sigma,j}-n_{-\sigma,j+1}\right) \\+\;& c_{\sigma,j}^\dagger c_{\sigma,j+1}^\dagger \left(n_{-\sigma,j}-n_{-\sigma,j+1}\right)\Big] +h.c.
\end{split}
\end{equation}

\begin{figure}[t]
	\centering
	\includegraphics[width=\linewidth]{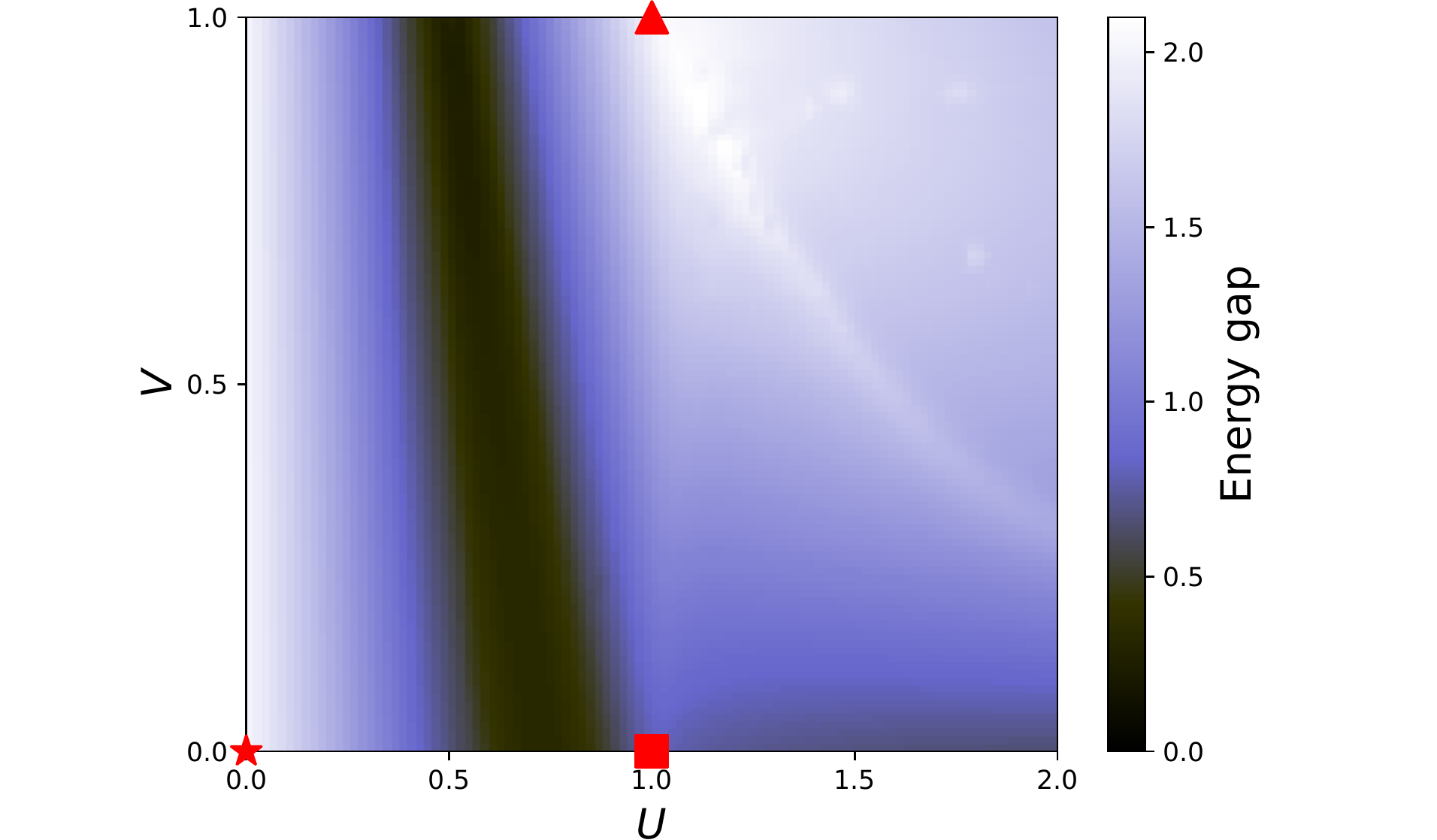}
	\caption{Energy gap [units $J$] of $\bar H(U,V)$  as a function of $U$ and $V$. Triangle, square and star correspond to $H_J$, $H_A$ and $H^{(2)}$ respectively. DMRG simulations on $16$ sites.  }
	\label{fig:phase}
\end{figure}

DMRG simulations on a chain with $16$ sites reveal a gap closure in the region $U \sim 0.5 - 0.7$, see Fig.\ \ref{fig:phase}. This defines two different phases: a ``strongly interacting'' (SI) one on the right and a ``weakly interacting'' (WI) one on the left. The original Hamiltonian $H_J$  [triangle in Fig.\ \ref{fig:phase}] belongs to the SI phase and can be continuously deformed into $H_A = \bar H (1,0)$ [square] -- an Hamiltonian in the $\mathbb{Z}_4$ parafermionic phase \textit{without} three-body interactions. {Note that, away from the exactly solvable point $H_J$, Hamiltonians $\bar H$ in the SI phase feature an exact four-fold degeneracy (through out all the spectrum) only in the $L\to\infty$ limit.}  

Our numerics thus show that parafermionic physics can already be generated from occupation-dependent hopping and pairing terms. Experimentally, such conditional terms can be realized if, e.g., the hopping involves intermediate virtual states whose energies are tuned by the interaction. Somewhat simpler density-dependent hoppings have already been engineered in cold-atomic systems \cite{jurgensen14,meinert16,Baier16}. On more general grounds, however, any not strictly local interaction gives rise to occupation-dependent hoppings and pairings \cite{kievelson1987,hirsch1989,hirsch1994}. It would be most desirable to identify (quasi-)one-dimensional systems in which these occupation-dependent terms are of appreciable size -- a challenging goal for future research that will also benefit from investigating the stability of the parafermionic phase under further modifications of the Hamiltonian.

{\it Fermionic spectral function.} From an experimental standpoint, a crucial (albeit not conclusive) signature of topological phases is the appearance of a zero-energy density of states at the ends of the topological chain. In general, the spin-averaged fermionic local spectral function at zero temperature reads
\begin{equation}
\begin{split}
A_j (\omega)&= 2\pi \sum_{\sigma,|\varphi\rangle} \Big[ \delta(\omega - E_{\varphi} +  E_{\rm GS}) \, | \langle \varphi | c_{\sigma,j}^\dagger |{\rm GS}\rangle|^2 \\
&\quad  + \delta(\omega + E_{\varphi} - E_{\rm GS}) \, \left| \langle \varphi | c_{\sigma,j} |{\rm GS}\rangle\right|^2  \Big]
\end{split}
\end{equation}
where $|\varphi\rangle$ are the eigenstates of the Hamiltonian with energies $E_\varphi$ and $|{\rm GS}\rangle$ is the ground state the system is in \footnote{For the numerical computation of the spectral function in presence of degeneracy, we select the ground state with odd fermion parity and with the lower expectation value of $M_1$. Different choices would not have modified the results we presented though.}. 

\begin{figure}[t]
	\centering
	\includegraphics[width=\linewidth]{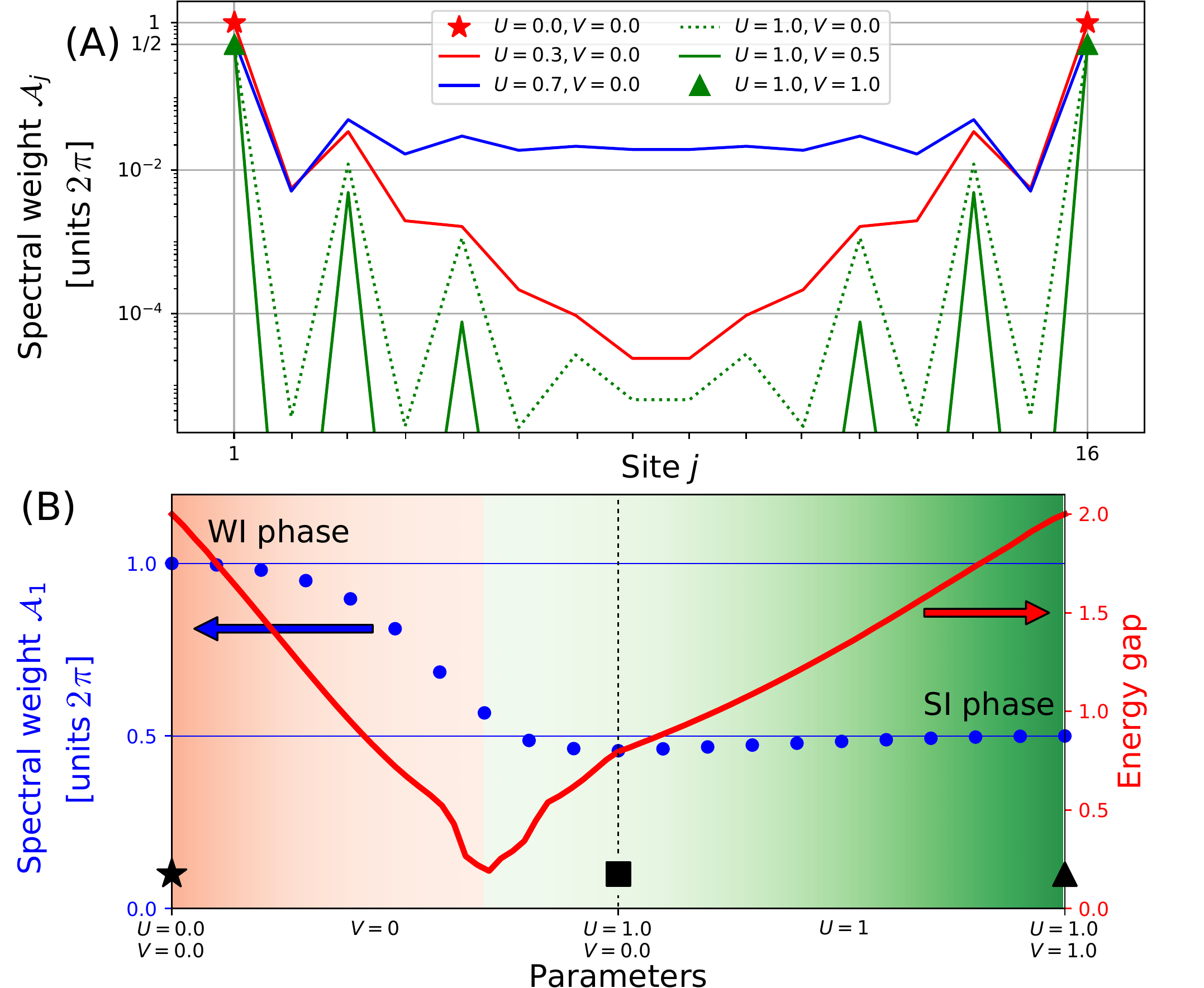}\\
	\caption{(A): 
	 $\mathcal{A}_j$ for different points in the $(U,V)$ parameter space. Green plots feature a dependence on the parity of the site $j$. (B): $\mathcal{A}_1$ (blue dots) along the straight paths in parameter space connecting $H^{(2)}$ [star], $H_A$ [square] and $H_J$ [triangle]. The energy gap [units $J$] is shown in red to help identifying the phase transition (here around $U\sim 0.7$) between the WI phase (red fade) and the SI one (green fade). DMRG simulations on $16$ sites. }
	\label{fig:spec}
\end{figure}

At first, we focus on the exactly solvable Hamiltonian $H_J$.  Denoting its four ground states with fixed FPF number $m$ (modulo $4$) by $|\psi_m \rangle$, one has that (see Appendix \ref{app:D})
\begin{equation}
\sum_m|\langle \psi_m|c_{j\sigma}|\psi_l \rangle |^2 = \left(\delta_{j,1}+\delta_{j,L}\right)\frac{1}{8} \qquad [H=H_{J}]
\end{equation}
for $l\in\{0,1,2,3\}$ and $\sigma= \uparrow,\downarrow$. The same holds true for the creation operators. Focusing on energies below the gap, this immediately leads to a zero-energy peak $A_{1,L}(\omega) = \pi\delta(\omega)$ localized at the edges and to a vanishing spectral weight in the bulk, $A_j(\omega) = 0$ for $j \in \{2, \ldots, L-1\}$. This result is confirmed by DMRG simulations which also allowed us to move away from the exactly solvable point. In particular, in Fig.\ \ref{fig:spec} we plot the spin-averaged spectral function integrated over the energy gap (EG) $\mathcal{A}_j = \int_{EG} A_j(\omega)\,  d\omega$ for different Hamiltonians.   Interestingly, the spectral weight is robust with respect to variations of the parameters $U$ and $V$ as long as the system remains in the SI phase. The fermionic edge density of state remains indeed trapped at the edges and features only an exponentially suppressed leakage into the bulk. This is clearly displayed in Fig.\ \ref{fig:spec} A. Note that the spectral weight within the gap has proven to be robust also with respect to other kind of small perturbations such as magnetic fields (along \emph{every} direction) and chemical potential. 

Fig.\ \ref{fig:phase} shows that a pronounced reduction of the interaction strength $U$ eventually leads to a phase transition, located where the gap closes (in a finite system the gap reaches a minimum but remains finite). At this point the low-energy spectral weight is spread all over the chain, as testified by the blue plot in Fig.\ \ref{fig:spec} A computed for $U=0.7$ and $V=0$. Once the system enters the WI phase, the spectral weight localizes again at the edges but with an important difference: as clearly shown in Fig.\ \ref{fig:spec} B, the low-energy spectral weight in the WI phase is twice the one in the SI one.  The reason is that the WI phase features two couples of zero-energy Majoranas instead of a single pair of parafermions. The non-interacting and exactly solvable Hamiltonian $H^{(2)}$ [red star in Fig.\ \ref{fig:phase}], which belongs to the WI phase, can indeed be expressed as two decoupled Kitaev chains with $4$ dangling edge Majoranas: 
\begin{equation}
H^{(2)} = -J i \sum_{j=1}^{L-1}\; \left[ \tau_{\downarrow,j}\, \chi_{\downarrow,j+1} \;+\; \tau_{\uparrow,j}\, \chi_{\uparrow,j+1} \right]
\end{equation}
where $\tau_{\sigma,j} = (\gamma_{-\sigma,j}+\eta_{\sigma,j})/\!\sqrt{2}\;$ and $\chi_{\sigma,j} = (\gamma_{\sigma,j}-\eta_{-\sigma,j})/\!\sqrt{2}$.
Moreover, it is possible to show that the four ground states of $H^{(2)}$ satisfy (see Appendix \ref{app:D})
\begin{equation}
\sum_m|\langle \phi_m|c_{j\sigma}|\phi_l \rangle |^2 = \left(\delta_{j,1}+\delta_{j,L}\right)\frac{1}{4} \qquad  [H=H^{(2)}],
\end{equation}
for $l\in\{0,1,2,3\}$ and $\sigma= \uparrow,\downarrow$. The same holds true for the creation operator. This leads to a peak $A_{1,L}(\omega) = 2\pi \delta(\omega)$ whose weight is exactly twice the one found in the SI phase. 

The zero-energy peak in the local spectral function, localized at the edges and with weight $\pi$ in a system with a time-reversal symmetric Hamiltonian provides therefore a robust signature of the SI phase and allows to distinguish between the presence of its $\mathbb{Z}_4$ parafermionic modes and the two couples of Majoranas featured by the WI phase. The existence of the phase transition between SI and WI underlines once more that inter-particle interactions play a crucial role for the emergence of zero-energy parafermions, as discussed also in Ref. \cite{pedder17,klinovaja13,klinovaja13b,vinkleraviv17}. 

{\it Discussion and conclusions.} In this Rapid Communication, we have introduced an exact mapping between $\mathbb{Z}_4$-parafermions and spinful fermions on a lattice. Despite the mapping's intrinsic non-locality, we showed that certain local parafermionic Hamiltonians (conserving the total number of Fock parafermions modulo $4$) can be mapped onto \emph{local} fermionic Hamiltonians. {This mapping thus allows for the systematic construction of interacting fermionic Hamiltonians on a lattice which feature zero-energy parafermionic modes.} 

In a first step, we focused on the exactly solvable fermionic model  $H_J$. The $\mathbb{Z}_4$ symmetry of the parafermionic model translates into the combination of $\mathbb{Z}_2$ fermion parity and a $\mathbb{Z}_2$ spontaneously broken symmetry. This challenges the topological protection of the zero-energy modes obeying $\mathbb{Z}_4$ parafermionic algebra, whose exact expressions in terms of fermions are derived. The lack of topological protection is in agreement with other very recent findings \cite{chew18,mazza18} on similar lattice systems. We studied experimentally accessible signatures of the $\mathbb{Z}_4$ parafermionic phase analytically, including in particular the fermionic density of states, and showed how it differs from other topological phases.

Importantly, the exactly solvable model $H_J$ belongs to an entire topological phase, which we explored in a second step. This phase includes much simpler Hamiltonians, which are more suitable for numerical and experimental investigation while featuring the same topological properties of $H_J$. 

Finally, the mapping we have introduced can be generalized to $\mathbb{Z}_{p}$ parafermions, which will be discussed elsewhere. In fact, as long as one chooses a suitable single-site basis, every local operator which conserves the total number of Fock parafermions modulo $p$ can be expressed in terms of fermions without string factors.

\begin{acknowledgments}
AC and TLS acknowledge financial support from the National Research Fund, Luxembourg, under grant ATTRACT 7556175. TM is supported by Deutsche Forschungsgemeinschaft through the Emmy Noether Programme ME 4844/1-1 and through SFB 1143. The authors thank Hosho Katsura for useful discussions. After the submission of this work, we became aware of Ref.\ \cite{chew18} that derives and studies a very similar model. The findings discussed in this reference are consistent with our results where they overlap.
\end{acknowledgments}

\appendix

\section{Mapping from parafermions to fermions}
\label{app:A}
Here we want to describe in details the procedure which allows to derive the full-lattice mapping between Fock parafermions (FPFs) and fermions, stated in Eq. (2) of the main text. 
First of all, it is worth noting that the parafermionic algebra of $a_i$ and $b_i$ is handed down to the FPFs in their properties: 
\begin{align}
\label{eq:d_lattice}
d_l d_j = i\; d_j d_l, \quad d^\dagger_l d_j = -i\; d_j d^\dagger_l& \quad \text{for } l<j,\\
\label{eq:d_onsite}
d_j^{\dagger m}d_j^m + d_j^{\dagger (4-m)}d_j^{4-m} = 1& \quad \text{for }m=1,2,3\\
\label{eq:A:d4}
d_j^4 = 0 &\quad \forall j.
\end{align}

As stated in the main text, the key idea of our mapping to electrons is to identify the single-site four-dimensional parafermionic Fock space with the Fock space of spin-$1/2$ fermions. We thus focus on a single site with FPF annihilation operator $d$ and fermionic operators $c_{\uparrow,\downarrow}$. A natural choice of a FPF basis is $\{|0\rangle, |1\rangle, |2\rangle, |3\rangle\}$, where $|n\rangle$ are the eigenstates of the FPF number operator $N$ with eigenvalue $n$. As for the fermions, we can for instance choose the basis 
\begin{equation}
\label{eq:BF}
\big\{ |E\rangle,\;  c^\dagger_{\uparrow}|E\rangle,\; i  c^\dagger_{\uparrow}  c^\dagger_{\downarrow}|E\rangle, \; -i  c^\dagger_{\downarrow}|E\rangle\big\},
\end{equation}
where $|E\rangle$ denotes the vacuum state, $c_{\sigma}|E\rangle=0$. The identification between these two bases induces the mapping 
\begin{equation}
\label{eq:mapping}
\begin{split}
d
&=  c^\dagger_{\downarrow}  c_{\uparrow}  c_{\downarrow} - c^\dagger_{\uparrow}  c_{\downarrow}^\dagger  c_{\downarrow}+ i  c^\dagger_{\uparrow}  c_{\uparrow}  c_{\downarrow} + c_{\uparrow},
\end{split}
\end{equation}
which automatically satisfies the algebra of Eq.~\eqref{eq:d_onsite} and Eq.\ \eqref{eq:A:d4}. In order to demonstrate the validity of the mapping in Eq.\ \eqref{eq:mapping}, we note that the action of the FPF annihilation operator $d$ on the basis states $|n\rangle$ is known by construction and reads $d |n\rangle = |n-1\rangle$ for $n=1,2,3$ and $d|0\rangle =0$. Its fermionic counterpart must behave in the same way when applied to the corresponding four fermionic basis states $|f_n\rangle$. Using the projectors on these states, one has
\begin{equation}
\label{eq:A:d}
\begin{split}
d& = 0 |f_0\rangle \langle f_0| + c_\uparrow|f_1\rangle \langle f_1| -i c_\uparrow |f_2\rangle \langle f_2| -c^\dagger_\uparrow |f_3\rangle \langle f_3|\\
& = c_\uparrow (1-n_\downarrow)(n_\uparrow) - i c_\uparrow (n_\uparrow n_\downarrow) - c^\dagger_\uparrow (1-n_\uparrow) n_\downarrow \\
& = c_\uparrow + c^\dagger_\downarrow c_\uparrow c_\downarrow + i  c^\dagger_\uparrow c_\uparrow c_\downarrow -  c^\dagger_\uparrow c^\dagger_\downarrow c_\downarrow,
\end{split}
\end{equation}
which corresponds to Eq.\ \eqref{eq:mapping}. This relation can be inverted, resulting in
\begin{align}
c_\uparrow & = d -d^\dagger d^2 - d^{\dagger 3} d^2,\\
c_\downarrow & = - i d^3 - i d^\dagger d^2 + i d^{\dagger 2} d^3\,.
\end{align}		
The next step is to extend this mapping over the full lattice. As pointed out in Ref.~[\onlinecite{cobanera14}], an FPF annihilation operator on site $j$ can be built as
\begin{equation}
\label{eq:A:d->dj}
d_j = i^{\sum_{p<j}N_p} \underbrace{\mathbb{I} \otimes \dots \otimes \mathbb{I}}_{j-1} \otimes\;  d \otimes  \underbrace{\mathbb{I} \otimes \dots \otimes \mathbb{I}}_{L-j}
\end{equation}
where $N_p = d_p^\dagger d_p + d^{\dagger 2}_pd_p^2 + d^{\dagger 3}_pd_p^3$ is the FPF number operator on site $p$ and the $\mathbb{I}$ are identity operators on a single site. This construction ensures that the FPF commutator algebra in Eq.\ \eqref{eq:d_lattice} is fulfilled. As for the fermions, one uses the standard Jordan-Wigner strings,
\begin{equation}
\label{eq:A:c->cj}
c_{j\sigma} = (-1)^{\sum_{\sigma}\sum_{p<j}n_{p\sigma}}  \underbrace{\mathbb{I} \otimes \dots \otimes \mathbb{I}}_{j-1} \otimes\;  c_\sigma \otimes  \underbrace{\mathbb{I} \otimes \dots \otimes \mathbb{I}}_{L-j}
\end{equation}
where $n_{p\sigma} = c^\dag_{p\sigma} c_{p\sigma}$ is the number of fermions with spin $\sigma$ on site $p$. Combining the above equations, one readily obtains the full lattice mapping
\begin{widetext}
	\begin{align}
	\label{eq:A:dj}
	d_j &= i^{\sum_{p<j} (N_p + 2n_{p\uparrow}+ 2 n_{p\downarrow})} \big( c^\dagger_{j\downarrow} c_{j\uparrow} c_{j\downarrow} + c_{j\uparrow} -c^\dagger_{j\uparrow} c_{j\downarrow}^\dagger c_{j\downarrow}  + i c^\dagger_{j\uparrow} c_{j\uparrow} c_{j\downarrow} \big)\\
	\label{eq:A:c_up}
	c_{j\uparrow} & = i^{\sum_{p<j} (-N_p + 2 n_{p\uparrow}+ 2 n_{p\downarrow})} \big[
	d_j - d_j^\dagger d_j^2 -(-1)^{\sum_{p<j} N_p} \; d_j^{\dagger 3} d_j^2 \big]\\
	\label{eq:A:c_down}
	c_{j\downarrow} & = i^{\sum_{p<j} (-N_p + 2 n_{p\uparrow}+ 2 n_{p\downarrow})}(-i)\big[
	(-1)^{\sum_{p<j} N_p}\;  d_j^3 + d_j^\dagger d_j^2 - d_j^{\dagger 2} d_j^3
	\big]\,.
	\end{align}
\end{widetext}

Interestingly, as stated in the main text, because of the definite odd fermion parity of Eq.\ \eqref{eq:A:dj} every parafermionic operators $D$ which conserves the number of FPFs modulo $4$ is transformed into a fermionic operator $C$ \emph{without} string factors. To see this, consider a generic operator $D$ which involves FPF operators on $m$ adjacent lattice sites
\begin{align}
D = d^{\dagger \alpha_0}_{j} d^{\beta_0}_j \; \dots\; d^{\dagger \alpha_m}_{j+m} d^{\beta_m}_{j+m}.
\end{align}
where $\alpha_k, \beta_k \in \{0, 1,2, 3\}$. Using the mapping in Eq.\ \eqref{eq:A:dj}, its fermionic expression factorizes into a form 
\begin{equation}
D = i^{\sum_{p<j} \left[\sum_{i=0}^m \left(\beta_i - \alpha_i\right)\right] \left(N_p + 2n_{p\uparrow} + 2 n_{p\downarrow}\right)}\;  C,
\end{equation}
with the operator $C$ containing, up to prefactors, only fermion operators on sites $j,$ $ \dots, j+m$. Requiring that $D$ conserves the total number of FPFs modulo $4$ then implies 
\begin{equation}
\sum_{i=0}^m \left(\beta_i - \alpha_i\right) = 0 \; (\text{mod } 4),
\end{equation}
so the string factor cancels. This remarkable result is at the heart of the locality of our mapping between parafermion chain Hamiltonians and electronic systems.\\

\section{Fermionic ground states}
\label{app:B}
In this section we focus on the properties of the four degenerate fermionic ground states of $H_J$. Following Ref.~[\onlinecite{iemini17}], the ground states $|\psi_m\rangle$ of $H_J$  (where $m = 0, \ldots, 3$ is the number of FPF modulo $4$) can be expressed as
\begin{equation}
\label{eq:GS}
|\psi_m\rangle = \frac{1}{\sqrt{4^{L-1}}} \sum_{\substack{
		\{n_j\} \text{ such that}\\\sum_j n_j = m\; (\text{mod } 4)}} \bigotimes_{j=1}^L |n_j\rangle
\end{equation}
where $|n_j\rangle$ is the single-site state with $n_j$ FPF. The proof of Eq.\ \eqref{eq:GS} is obtained by introducing the operator
\begin{equation}
\xi_j = \frac{1}{\sqrt{2}} \left[ d_j i^{N_j} + d^{\dagger 3}_j - i \left(d_{j+1}+d^{\dagger 3}_{j+1}\right)\right]
\end{equation}
which allows to write the Hamiltonian as 
\begin{equation}
H_J = - 2 J (L-1) + 2J \sum_{j=1}^{L-1} \xi_j^\dagger \xi_j\,,
\end{equation}
where the second term is non-negative. A state is a ground state if $\xi_j |\psi \rangle = 0\; \forall j$, a condition which is fulfilled by Eq.\ \eqref{eq:GS}. Note that excited states can be obtained from the ground states $|\psi_m \rangle$  by applying $\xi_j^\dagger$. The energy gap between the ground states and the first excited states is $2 J$. 

Our mapping allows us to express these ground states in terms of fermions: one simply has to replace each FPF single-site state $|n_j\rangle$ with the corresponding fermionic one, which we will call $|f_{n_j}\rangle$. This will allow us to directly study some interesting fermionic properties of the ground states. In the following, the integer indexes of $|\psi_m\rangle$ and $|f_n\rangle$ are always understood modulo $4$.  

\subsection{Fermionic edge operators}
The goal here is to show that, despite the intrinsic non-locality of $b_L$ (see Eq.~(7) in the main text), it is still possible to find \emph{local} operators at either edge of the fermionic chain that cycle through the four degenerate ground states $|\psi_m\rangle$.

If we focus on the left edge ($j=1$), it is easy to show that the parafermionic operator $a_1$ actually does the job. Indeed, being on the first site of the chain, its fermionic expression has no string factors (see Eq.~(7) in the main text). Moreover, it decreases by one the number of FPF modulo $4$ on the first chain site. Therefore, one has for all ground states,
\begin{equation}
a_1 |\psi_{m} \rangle  = | \psi_{m-1 }\rangle 
\end{equation}
Things are more complicated at the right edge ($j=L$), where one has to take care of string factors. For notational convenience, let us introduce the following state of a fermionic chain with $S$ sites
\begin{equation}
|b^S_m\rangle = \frac{1}{\sqrt{4^{S-1}}} \sum_{\substack{\{n_j\} \text{ such that}\\\sum_j n_j = m\; (\text{mod } 4)}} \bigotimes_{j=1}^{S} |f_{n_j}\rangle\,.
\end{equation}
This allows us to express the four ground states as
\begin{align}
\label{eq:A:psi_L}
|\psi_m\rangle &= \frac{1}{2} \Big(|b^{L-1}_0 \rangle \otimes |f_m\rangle+|b^{L-1}_1 \rangle \otimes |f_{m-1}\rangle \notag \\
&+|b^{L-1}_2 \rangle \otimes |f_{m-2}\rangle+|b^{L-1}_3 \rangle \otimes |f_{m-3}\rangle\Big) 
\end{align}
In view of the chosen fermionic basis (see Eq.~\eqref{eq:BF}), the state $|b_m^{L-1}\rangle$ contains an even or an odd number of fermionic operators acting on the empty chain, depending on whether $m$ is even or odd, respectively. As a consequence, in order to be able to cycle between the four ground states, a local fermionic edge operator $\gamma_L$ must satisfy
\begin{equation}
\gamma_L \Bigg(\bigotimes_{j=1}^{L-1} \mathbb{I} \Bigg) \otimes |f_m\rangle = (-1)^{m-1}  \Bigg(\bigotimes_{j=1}^{L-1} \mathbb{I} \Bigg) \otimes |f_{m-1}\rangle\,.
\end{equation}
It is easy to show that
\begin{equation}
\gamma_L =  c^\dagger_{L\uparrow}n_{L\downarrow}-i c_{L\downarrow} n_{L\uparrow}+c_{L\uparrow}(1-n_{L\downarrow})+ic^\dagger_{L\downarrow}(1-n_{L\uparrow})
\end{equation}
satisfies the above condition and thus represents a fermionic edge operator which cycles between the four ground states
\begin{equation}
\gamma_L |\psi_{m} \rangle = |\psi_{m-1} \rangle
\end{equation}
{As discussed in Section \ref{sec:H2}, this feature is peculiar of $H_J$ (and the SI phase): in the WI one it is not possible to find such operators.}
\section{Effects of local fermionic perturbations}
\label{app:C}
\begin{figure}
	\centering
	\includegraphics[width=\columnwidth]{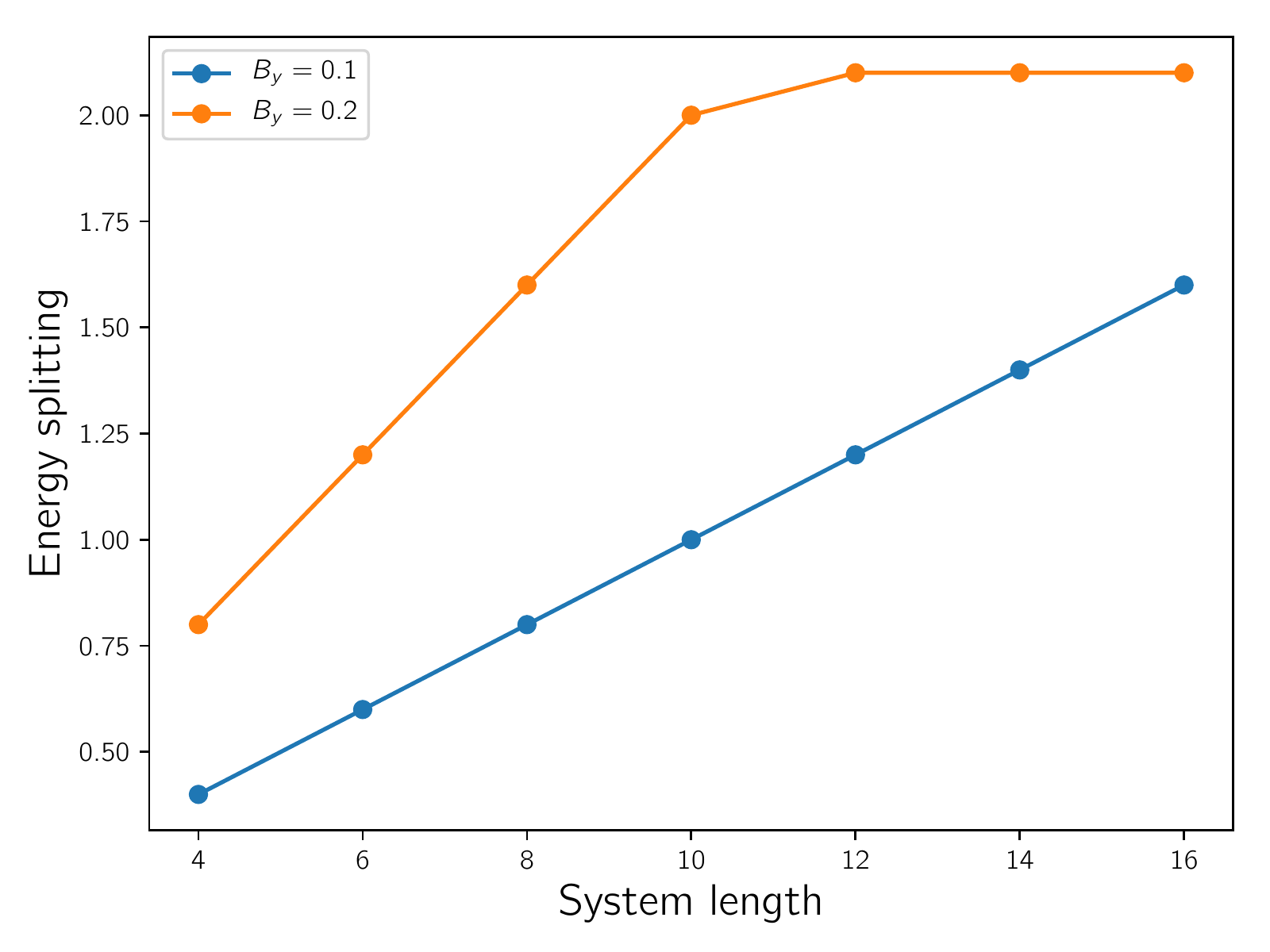}
	\caption{Lifting of the ground state degeneracy $\Delta E_y$ [units $J$] for Hamiltonian $H_J+H_y$, as a function of system length $L$ for different values of $B_y$ [units $J$].}
	\label{fig:1}
\end{figure}

Here we compute, using DMRG simulations, the effects of some local fermionic perturbations on the $4$-fold degeneracy of the ground states. At first, we focus on a magnetic field along the $y$ axis (see Eq. (8) in the main text)
\begin{equation}
H_{y} = B_y \sum_{j=1}^L i \left(c^\dagger_{j,\uparrow}c_{j,\downarrow}-c^\dagger_{j,\downarrow}c_{j,\uparrow}\right)=B_y \sum_{i=1}^L \tfrac{1}{2}\left(M_j + i \eta_{\uparrow,j}\eta_{\downarrow,j}\right)
\end{equation}
which explicitly contains the local order parameter $M_j$ and it is therefore expected to split the $4$ ground states into two doublets. They are separated by energy $\Delta E_y$ which grows linearly with the system size, as shown in Fig.\ \ref{fig:1}. The saturation around $\Delta E_y \sim 2 J$ is due to the presence of the bulk gap. 

\begin{figure}
	\centering
	\includegraphics[width=\columnwidth]{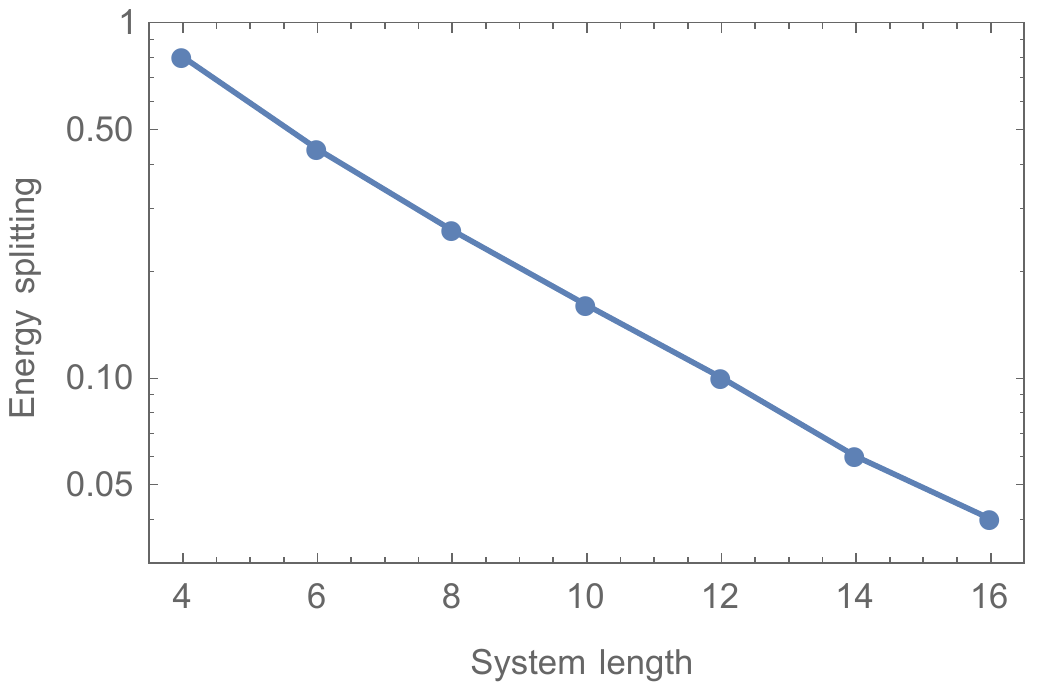}
	\caption{Liftings of the ground state degeneracy $\Delta E_\mu$ [units $J$] as a function of system length $L$. Here we choose $\mu=1.6 J$}
	\label{fig:2}
\end{figure}

By contrast, magnetic fields in the $(x,z)$-plane and chemical potential terms
\begin{align}
H_{z} &=  B_z \sum_{j,\sigma} \sigma  \; c^\dagger_{\sigma,j} c_{\sigma,j} 
\\H_{x} & =  B_x \sum_{j} c^\dagger_{j,\downarrow}c_{j,\uparrow} + h.c. 
\\H_{\mu} &=  \mu \sum_{j,\sigma}  c^\dagger_{\sigma,j} c_{\sigma,j} 
\end{align}
lead to a degeneracy lifting which is exponentially suppressed with the system length $L$. This is clearly shown in Fig.\ \ref{fig:2} where we display the energy splitting $\Delta E_\mu$ between the ground state and the fourth eigenstate of the system when a chemical potential term $H_\mu$ is added to $H_J$. The energy splittings $\Delta E_z$ and $\Delta E_x$ (due to $H_z$ and $H_x$ respectively) happen to be the same as $\Delta E_\mu$. 

Interestingly, both $H_\mu$ and $H_z$, conserve the total number of FPF modulo $4$ and they thus feature a local expression also in terms of parafermions: neglecting an inessential constant term, one has indeed
\begin{equation}
\begin{split}
H_f = - \sum_{j,\sigma} (\mu+ \sigma B_z)\,  c^\dagger_{\sigma,j} c_{\sigma,j} \sim \sum_j \!e^{i \frac{\pi}{4}} \frac{B_z-i\mu}{2} a^\dagger_j b_j + h.c.\,. 
\end{split}
\end{equation}
Therefore it is clear that they cannot lift the four-fold degeneracy in an infinite system. Moreover, note that the $H = H_J+ H_f$ correspond to well known nearest-neighbor four-state clock model  \cite{fendley12,fendley14,alicea16} which features full topological protection for small $B_z$ and $\mu$. 

\section{Fermionic spectral function}
\label{app:D}
In this section we are going to compute analytically the fermionic spectral function for the two exactly solvable models presented in the main text: $H_J$ [see Eq. (4-6)] which belongs to the SI phase and $H^{(2)}$ [see Eq. (4)] which belongs to the WI phase.

\subsection{Hamiltonian $H_J$}
Let us consider the Hamiltonian $H_J$ which contains exact parafermionic zero-energy modes. Here we will prove that, despite the presence of string factors in the expression of $b_L$, the zero-energy fermionic spectral function is non-vanishing only at the edges of the chain, where it has the universal prefactor $1/4$. To this end, we observe that the only non-vanishing matrix elements of the fermionic annihilation operators on a single site are
\begin{align}
\langle f_0 | c_\uparrow | f_1 \rangle = 1\,,  \qquad  \langle f_3 | c_\uparrow | f_2 \rangle = -1\,,\notag \\
\langle f_1 | c_\downarrow | f_2 \rangle = -i\,,  \qquad  \langle f_0 | c_\downarrow | f_3 \rangle = -i\,.
\end{align}
This, together with the expression for the ground states in Eq.~\eqref{eq:A:psi_L}, allows to prove by direct calculation
\begin{align}
|\langle \psi_{m\pm 1} |c_{L\sigma}|\psi_m \rangle|^2 &= \frac{1}{16}\,,\notag \\
\qquad |\langle \psi_m  |c_{L\sigma}|\psi_m \rangle|^2 &= |\langle \psi_{m+2}  |c_{L\sigma}|\psi_m \rangle|^2= 0\,.
\end{align}
The same argument applies for the left edge of the chain ($j=1$), since the ground states can be equivalently expressed also as
\begin{align}
|\psi_m\rangle &= \frac{1}{2} \Big( |f_m\rangle \otimes |b^{L-1}_0 \rangle +|f_{m-1}\rangle \otimes |b^{L-1}_1 \rangle \notag \\
&+ |f_{m-2}\rangle \otimes |b^{L-1}_2 \rangle + |f_{m-3}\rangle \otimes |b^{L-1}_3 \rangle \Big)\,.
\end{align}
This leads us to
\begin{equation}
\sum_{l} |\langle \psi_l | c_{j\sigma} | \psi_m \rangle|^2 = \frac{1}{8}\qquad \text{for } j=1,L.
\end{equation}
Let us now consider a bulk site $k \notin \{1, L\}$ and prove that the matrix elements of $c_{k\sigma}$ between the ground states are zero. The key observation is that $|\psi_m\rangle$ can also be expressed as
\begin{align}
|\psi_m\rangle &= \frac{1}{4}  \sum_{a=0}^3 \Big( |b^{k-1}_a\rangle \otimes |f_m\rangle \otimes |b^{L-k}_{-a} \rangle + |b^{k-1}_a\rangle \otimes |f_{m-1}\rangle \otimes |b^{L-k}_{1-a} \rangle\notag \\ &+  |b^{k-1}_a\rangle \otimes |f_{m-2}\rangle \otimes |b^{L-k}_{2-a} \rangle +|b^{k-1}_a\rangle \otimes |f_{m-3}\rangle \otimes |b^{L-k}_{3-a} \rangle \Big)\,.
\end{align}
Since $a$ is not fixed, the matrix elements reads
\begin{align}
\langle \psi_{m\pm 1}|c_{k\sigma}|\psi_m\rangle &\propto \sum_{a=0}^3 (-1)^a =0\,,\notag \\
|\langle \psi_m  |c_{k\sigma}|\psi_m \rangle|^2 &= |\langle \psi_{m+2}  |c_{k\sigma}|\psi_m \rangle|^2= 0\,
\end{align}
with the factor $(-1)^a$ stemming from the anticommutation of $c_{k\sigma}$ with the fermionic operators contained in $|b^{k-1}_a\rangle$. Since the above results hold also for the fermionic creation operators $c^\dagger_{j\sigma}$, the properties of the spectral function discussed in the main text are proved.


\subsection{Hamiltonian $H^{(2)}$}
\label{sec:H2}
Here we focus on the non-interacting Hamiltonian $H^{(2)}$ which consists of two uncoupled Majorana chains [see Eq. (19)]
\begin{equation}
H^{(2)} = -J i \sum_{j=1}^{L-1}\; \left[ \tau_{\downarrow,j}\, \chi_{\downarrow,j+1} \;+\; \tau_{\uparrow,j}\, \chi_{\uparrow,j+1} \right]
\end{equation}
with
\begin{align}
\label{eq:A:chi_sigma}
\chi_{\sigma,j} &= \frac{1}{\sqrt{2}} \left[c_{\sigma,j}^\dagger +c_{\sigma,j}+i \left(- c_{-\sigma,j}^\dagger+ c_{-\sigma,j}\right)\right]  \\
\tau_{\sigma,j} &= \frac{1}{\sqrt{2}} \left[ c_{-\sigma,j}^\dagger + c_{-\sigma,j} + i \left(c_{\sigma,j}^\dagger-c_{\sigma,j}\right)\right]\,.
\end{align}
It is possible to define two non-local fermionic operators
\begin{equation}
\label{eq:A:f}
\mathtt{f}_\sigma = \frac{1}{2} \left(\tau_{\sigma,L} + i \chi_{\sigma,1}\right)
\end{equation} which commute with $H^{(2)}$ and allows to label the four ground states of the system
\begin{equation}
\label{eq:A:H2_gs}
|\phi_0\rangle, \; |\phi_1\rangle = \mathtt{f}^\dagger_{\downarrow} |\phi_0\rangle, \; |\phi_2\rangle = \mathtt{f}^\dagger_{\uparrow} \mathtt{f}^\dagger_{\downarrow} |\phi_0\rangle, \; |\phi_3\rangle =\mathtt{f}^\dagger_{\uparrow} |\phi_0\rangle
\end{equation}
where here $|\phi_0\rangle$ is the ground state of $H^{(2)}$ annihilated by both $\mathtt{f}_\sigma$ ($\sigma=\uparrow,\downarrow$). Given the expression in Eq.\ \eqref{eq:A:chi_sigma}-\eqref{eq:A:H2_gs}, one can immediately show that 
\begin{equation}
\sum_{m=0}^3 |\langle \phi_m| c_{j,\sigma}^\dagger |\phi_n\rangle|^2 = \sum_{m=0}^3 |\langle \phi_m| c_{j,\sigma} |\phi_n\rangle|^2 = (\delta_{j,1}+\delta_{j,L})\; \frac{1}{4} 
\end{equation} 
for all $n=0,1,2,3$ and $\sigma$, thus proving Eq. (14) in the main text. 

{We conclude the discussion about the WI phase and its exactly solvable model $H^{(2)}$ by underlying that, differently from the parafermionic phase, here it is not possible to find a single edge operator which cycle between the four ground states. Let us focus on the left edge, for example. Here the most generic operator $\Xi_1$ which commutes with the Hamiltonian $H^{(2)}$ is a combination of $\chi_{\uparrow,1}$ and $\chi_{\downarrow,1}$
	\begin{equation}
	\Xi_1 = \alpha \chi_{\downarrow,1}\ + \beta \chi_{\uparrow,1}\,.
	\end{equation}
	Being the latter two Majorana operators, $\Xi_1^2$ is necessarily proportional to the identity 
	\begin{equation}
	\Xi_1^2 = \alpha^2+\beta^2 + \alpha\beta \; \{\chi_{\uparrow,1},\chi_{\downarrow,1}\} =\alpha^2+\beta^2
	\end{equation}
	and it can therefore cycle only between two different states.}

\bibliography{refs}

\end{document}